\documentclass{iau}

\usepackage{amsmath}
\usepackage{graphicx}
\usepackage{multirow}
\usepackage{xspace}
\def\gaia{\textit{Gaia}\xspace}

\begin{document}

\lefttitle{Eyer, Huijse, Chornay, et al.}
\righttitle{Proceedings of the International Astronomical Union}

\jnlPage{1}{7}
\jnlDoiYr{2021}
\doival{10.1017/xxxxx}

\aopheadtitle{Exploring the Universe with Artificial Intelligence (UniversAI)}
\editors{I. Liodakis, eds.}

\title{The “Variable” Universe 
with the \gaia mission and AI methods}

\author{L. Eyer$^1$, P. Huijse$^2$, N. Chornay$^3$, J. De Ridder$^2$, B. Holl$^{1,3}$, L. Rimoldini$^3$, K. Nienartowicz$^{3,4}$, G. Jevardat de Fombelle$^{3}$}
\affiliation{
$^{1 }$Department of Astronomy, University of Geneva, Chemin Pegasi 51, 1290 Versoix, Switzerland\\
$^{2 }$Institute of Astronomy, KU Leuven, Celestijnenlaan 200D, 3001 Leuven, Belgium\\
$^{3 }$Department of Astronomy, University of Geneva, Chemin d'Ecogia 16, 1290 Versoix, Switzerland\\
$^{4 }$Sednai sàrl, 4 Rue de Marbiers, 1204, Geneva, Switzerland}

\begin{abstract}
The \gaia mission has observed over 2 billion stars repeatedly across the entire sky over 10\,years, revealing the many astronomical objects that vary on human timescales from seconds to years. Its repeated astrometric, photometric, spectrophotometric and spectroscopic measurements create an unprecedented dataset to probe the variable celestial sources down to $G\approx 21$\,mag. To extract meaningful results from these many time series for so many sources, we have used machine learning techniques for crossmatching, variability detection, and variability classification. This approach has now led to the largest catalogue of classified variable sources ever produced over the entire celestial sphere.
\end{abstract}


\maketitle

\section{Introduction}
Time-domain astronomy is an extremely rich and diverse field, yet we can still be amazed that so many celestial phenomena reveal themselves on human time scales. 
Among the many time-domain projects, the \gaia mission \citep{GCPrustiEtal2016} provides remarkable insights into the "variable sky". \gaia is a cornerstone mission of the Horizon 2000+ program of the European Space Agency. The spacecraft was launched in December 2013 and was passivated in March 2025.
\gaia is unique in that, on a single platform and nearly simultaneously, it collected astrometric, photometric, spectrophotometric, and spectroscopic data across the entire celestial sphere, over a period of more than 10\,years. Furthermore, thanks to its astrometric time series, the celestial sphere gets a geometrical depth!

So far, there have been three major data releases: DR1 in 2016 \citep{GCBrownEtal2016}, DR2 in 2018 \citep{GCBrownEtal2018}, and EDR3/DR3 in 2020/2022 \citep{GCBrownEtal2021, GCVallenariEtal2023}. In addition, a focused release in 2023 included five articles, one of which was dedicated to the radial velocities of long-period variables \citep{GCTrabucchiEtal2023}. Two further releases are planned: DR4 at the end of 2026 and DR5 in 2030.

\gaia has been an enormous success: its data are used across  so many fields of astrophysics, and its publications break citation records \citep{ParmarEtal2024}.

The vast number of sources (more than 2 billion) and the diversity of variable phenomena make automated machine-learning approaches indispensable. In this article, we focus specifically on the use of machine-learning methods for classifying variability in \gaia time-series data. The processing of \gaia data is iterative: with each new release, the data volume grows, calibrations improve, and the outputs become more diverse and complex. As a result, the development of methods and software has been a continuous effort.

The understaking to classify variable sources within \gaia has its roots in the \textit{Hipparcos} mission \citep{ESA1997, Eyer1998, WaelkensEtal1998, AertsEtal1998}, with a mixture of procedural approach rooted in astrophysical knowledge and also a multivariate discriminant analysis.  Exploratory work on clustering techniques was carried out by \cite{EyerBlake2005} on ASAS data, using k-means applied to fundamental variability features such as the period, amplitude, and the parameters of a Fourier series. Other methods were explored for \gaia; e.g. the stripe 82 of SDSS multiband photometry was analysed with a principle component analysis \citep{SuvegesEtal2012}.
But a crucial step was achieved when the \textit{Hipparcos} dataset was used to design classification methods for \gaia. Random Forest algorithms \citep{Breiman2001} proved particularly effective \citep{DubathEtal2011}. In this latter article, a comparison with a hierarchical classification approach developed by \cite{BlommeEtal2011} for periodic variables was made and confirmed the effectiveness of the Random Forest approach. \cite{RimoldiniEtal2012} applied Random Forest method to the "unsolved" variable stars of the \textit{Hipparcos} mission and showed again its pertinence.
Today, \gaia itself serves as a sandbox for developing and testing new methods. The latest efforts include the use of autoencoders (see Section~\ref{Sect:Autoencoder}).
A particular feature of \gaia processing is that variability classification is first performed globally, after which specific classes are passed on for more detailed studies: for example, Cepheids and RR\,Lyrae stars. This creates a feedback loop between the machine-learning classifications and the evaluations of expert astronomers. Furthermore, for all classes, efforts were made to estimate completeness and contamination.

\section{Supervised Classification}
A central task of our activity in the \gaia Data Processing and Analysis Consortium  \citep[DPAC, ][]{MignardEtal2008} is given the \gaia time domain data to classify the source into variability types.
Supervised classification was applied to \gaia Data from the first data release, then the second and the third.
The method applied, as mentioned above, was based on feature extraction followed by the use of random forest classifiers. A total of 28 features were selected (from a much larger set) in the classification, some derived from periodograms and others from statistical measures; the full list is provided in \cite{RimoldiniEtal2023}. In addition, we compiled data from the literature to construct the training sets. As the adage goes, in supervised classification: garbage in, garbage out. The quality of the training set is therefore crucial. Thanks to the \gaia data, we can investigate variable sources reported in the literature. Typical problems with incorrectly labelled data include a variability level inconsistent with \gaia measurements, a light-curve shape incompatible with the assigned variability type, implausible \gaia colours and, when parallax is available, an incorrect position in the Hertzsprung–Russell diagram. The analysis of the published literature has often been of varying quality, with some notable positive exceptions, most prominently the OGLE project \cite{UdalskiEtal1992}.

At every data release, the number of classified sources and the number of variability types are larger.
DR1 contained 3,194 sources in two classes (Cepheids and RR Lyrae classes) for a limited region in the sky (part of the Large Magellanic Cloud). This release was more a showcase, a teaser for what is to come \citep{EyerEtal2017}.
DR2 had 550,737 variable stars classified in 9 variability (sub)types \citep{HollEtal2018, RimoldiniEtal2019}.
DR3: listed 10.5 million variables in total into 24 \footnote{In fact, there are 24+1 variable types, because the supervised classification was able to identify 2.5 million galaxies due to spurious variability caused by the extended nature of these objects and the way Gaia acquired these data.} variability (sub)types, including 9.5 million variable stars and 1 million variable AGNs. Moreover, \gaia has made these classifications possible with nearly simultaneous measurements in astrometry, photometric time series, spectrophotometry, spectroscopy, and radial velocities (the last two for brighter sources) \citep{EyerEtal2023a,RimoldiniEtal2023}. The photometric time series are available, the spectrophotometry was used to classify carbon stars and an unprecedented sample of radial velocities time series was published
\citep{LebzelterEtal2023}.
The photometric and radial velocity time series are publicly available in the \gaia ESA archive\footnote{ \url{https://gea.esac.esa.int/archive/}}.

One characteristic of working within an ESA consortium is that we are bound to a tight schedule. Processing time is often a bottleneck. Although, as Donald Knuth famously remarked, “premature optimisation is the root of all evil”, we sometimes need to make non-optimal choices to reduce the time required to meet deadlines. In recent years, a major game changer has been the use of GPUs \citep{Fluke2012}, which have significantly accelerated many time-consuming computations, such as period-search methods.

\section{Variational Autoencoders}
\label{Sect:Autoencoder}
Autoencoders are neural network methods that reduce the dimensionality of input data  \citep{goodfellow2016deep}. An encoder compresses the input (for example, an image or a time series) into a lower-dimensional representation, and a decoder reconstructs the input from this compressed form. 
By training the network to minimise the difference between the reconstruction and the original, the autoencoder learns a reduced but meaningful description of the data. The compressed quantities are referred to as the latent variables.

For DR4, we will provide a separate catalogue output based on variational autoencoders (VAEs), building on work carried out with \gaia DR3 data \citep{HuijseEtal2025}. In that study, three VAEs were trained, each with five latent variables, on different \gaia data products (1) the BP and RP mean spectrophotometry, (2) the folded G light curve,  and (3) the DMDT (magnitude differences versus the corresponding time pairs) representation of the light curve \citep{MahabalEtal2017}. The folded light curve is especially informative for periodic sources, while the DMDT method better captures aperiodic variability. 
From these three VAEs, 15 latent variables were obtained. After normalising the inputs, three additional features were added: the log of the dominant frequency, the standard deviation and the colour. Together, these 18 features provide a compact yet physically motivated description of each source. Importantly, parallax (and thus distance) was not included.

A second-stage VAE was then applied to reduce these features to two latent variables, yielding a 2D projection of the data. This projection successfully groups the different major classes and shows strong correlation with astrophysical properties, demonstrating the method's effectiveness.
\citet{HuijseEtal2025} also showed that the latent representation can serve as input to other machine learning methods, enabling:
\begin{itemize}
\item Anomaly detection, when combined with techniques such as local outlier factor or isolation forests.
\item Clustering, using methods such as $k$-means or DBSCAN \citep[Density-Based Spatial Clustering of Applications with Noise,][]{EsterEtal1996}.
\item Classification, in combination with algorithms such as $k$-NN or random forest.
\end{itemize}
The forthcoming DR4 analysis will allow us to fully assess the potential of this approach.

\section{Classification with GaiaVari, a citizen science project}
GaiaVari\footnote{\url{https://www.gaiavari.space}} is a citizen science project on the Zooniverse platform aimed at classifying subsets of variable sources from \gaia DR3 \citep{EyerEtal2023b, Eyer2024} 
The classification relies on fundamental astronomical plots, such as time series, folded light curves, HR/colour–magnitude diagrams, and location in the Milky Way. We believe it is essential to promote high-level interaction not only between machine-learning experts and astronomers but also with the wider public. Such an activity demonstrates the fundamental importance of visualising properties of individual sources. This does not mean that having many sources makes it pointless to look at just a few. On the contrary, the more sources you have, the more meaningful subsamples you can create!

\section{Conclusion and perspective}
For the variable sources, we have been preparing the data processing and analysis of the \gaia mission for many years within the \gaia DPAC Consortium, testing and applying different machine-learning methods. Our goal has been to produce catalogues with a holistic approach that can be used by the scientific community and amateur astronomers. The result is a significant leap forward: at the time of the third data release publication, we provided, across the entire sky, the largest and most homogeneous set of variable sources to date, breaking many records in the number of many variability types. Our analysis provides, for the first time, a comprehensive view of variability among Milky Way stars across the Colour–Absolute Magnitude Diagram \cite{GCEyerEtal2019}.

This is not the end of the road for the \gaia data and associated methods to analyse them. Although the spacecraft terminated its data collection, future releases will include iteratively larger datasets, and machine learning techniques are advancing rapidly. In DR4, the time series used for analysis will be twice as long as in DR3, and it will double again in DR5. Also note that from DR4 all time series will be available to the public. From other surveys, we will be able to benefit from OGLE, ZTF and TESS data for the training sets and validation.

Within the \gaia consortium, each Data Release is an opportunity to test new methods. Indeed, \gaia provides a valuable testbed for novel machine-learning approaches, with the published catalogues serving as benchmarks against which new techniques can be compared.

In time-domain astronomy, the next major leap forward will likely come from the Vera Rubin Observatory (LSST; \citealt{LSST2009}), which will explore the faint variable sky from the ground using a multi-band photometric system. Combining the strengths of both \gaia and Rubin will enable significant progress in the field over the coming years.

\end{document}